\documentclass[shortnote,twocolumn]{jpsj3}

\title{
Lower Critical Field $H_{\rm c1}(T)$ and Pairing Symmetry 
Based on Eilenberger Theory
}

\author{
Takanobu Akiyama, 
Masanori Ichioka, 
and Kazushige Machida
} %\\
\inst{
Department of Physics, Okayama University,
Okayama 700-8530, Japan
}
\abst{
We quantitatively estimated different T-dependences of Hc1 between s wave and d wave pairings by Eilenberger theory. The T-dependences of Hc1(T) show quantitative deviation from those in London theory. We also studied differences of Hc1(T) between p+ and p- wave pairing in chiral p wave superconductors. There, Hc1(T) is lower in p- wave pairing, and shows the same T-dependence as in s wave pairing. 
}

\kword{
Vortex states, 
Lower critical field, 
Eilenberger theory, 
Anisotropic superconductors, 
d-wave pairing, 
Chiral p-wave pairing}

\begin{document}
\maketitle

\def\runtitle{
Lower critical field $H_{\rm c1}(T)$ and pairing symmetry 
based on Eilenberger theory
}
\def\runauthor{
%T.Akiyama, {\it et al.} 
T. Akiyama, 
M. Ichioka, 
and K. Machida
}
%%%%%%%%%%%%%%%%%%%%%%%%%%%%
%\section{Introduction}

Vortex physics plays an important role in the study of 
unconventional superconductors. 
In this short note, 
based on Eilenberger theory 
we study the temperature ($T$) dependence of lower critical field 
$H_{\rm c1}(T)$ of vortex states in anisotropic superconductors.

Recent developments of experimental technique make us possible to 
observe $H_{c1}(T)$ 
exactly, and discuss it 
in the relation to the mechanism of unconventional superconductivity, 
such as in new iron-based superconductors~\cite{Okazaki,Pribulova} 
and ${\rm URu_2Si_2}$.\cite{URu2Si2} 
In traditional Ginzburg-Landau (GL) theory, 
$H_{\rm c1}$ is given by~\cite{Fetter}   
\begin{eqnarray}
H_{\rm c1} \propto \lambda(T)^{-2} ( \ln \kappa + c_0)
\label{eq:hc1gl}
\end{eqnarray}
with penetration depth $\lambda$,  GL parameter $\kappa$, 
and a constant $c_0$. 
In the London theory 
$\lambda(T)=\lambda(T=0) +\delta \lambda(T)$ behaves as 
$\delta \lambda (T)  \propto  \exp(-\Delta/k_{\rm B}T)$ 
at low $T$ in the $s$-wave pairing, 
reflecting superconducting gap $\Delta$. 
In $d$-wave pairing with line nodes, 
$\delta \lambda(T) \propto T$ at low $T$ in the clean limit. 
These indicate that $H_{\rm c1}(T)$ depends on  
the pairing symmetry of anisotropic superconductors. 

We note that GL theory is a phenomenological theory valid near 
the transition temperature $T_{\rm c}$. 
Thus it is not clear whether  
the above-discussion on $H_{\rm c1}(T)$ is quantitatively valid. 
Therefore, it is expected that $H_{\rm c1}(T)$ is 
evaluated by Eilenberger theory, which is 
quantitatively reliable in vortex states even far from $T_{c}$.   
To study contributions by the pairing symmetry, 
we calculate $H_{\rm c1}(T)$ for $s$-wave pairing with full gap
and $d_{x^2-y^2}$-wave pairing with line nodes, as typical examples, 
by quantitative Eilenberger theory.  
The previous work for $s$-wave pairing was done 
in a single vortex.~\cite{KPssc} 
Our calculation is performed in vortex lattice. 
We also study $H_{\rm c1}(T)$ for chiral $p_\pm$-wave pairings, 
to see dependences on the chirality directions, i.e., 
parallel or anti-parallel to applied fields. 

%%%%%%%%%%%%%%%%%%%%%%%%%%%%
%\section{Formulation}

In this study, for simplicity, 
we use isotropic cylindrical Fermi surface 
${\bf k}=k_{\rm F}(\cos\theta,\sin\theta)$ and magnetic fields 
are applied to the $c$ direction.    
The quasiclassical Green's functions
$g( \omega_n, {\bf k},{\bf r})$, $f( \omega_n, {\bf k},{\bf r})$, and 
$f^\dagger( \omega_n, {\bf k},{\bf r})$  
are calculated by the Eilenberger equation 
\begin{eqnarray} &&
\left\{ \omega_n 
+{\bf v} \cdot\left(\nabla+{\rm i}{\bf A} \right)\right\} f
=\Delta \phi g, 
\nonumber 
%\label{eq:eil1}
\\ && 
\left\{ \omega_n 
-{\bf v} \cdot\left( \nabla-{\rm i}{\bf A} \right)\right\} f^\dagger
=\Delta^\ast \phi^\ast g  , \quad 
%\label{eq:eil2}
\label{eq:eil}
\end{eqnarray} 
$g=(1-ff^\dagger)^{1/2}$ 
%${\rm Re} g > 0$, 
in the vortex lattice state, 
with the selfconsistent conditions of pair potential  
\begin{eqnarray}
\Delta({\bf r})
= g_0N_0 T \sum_{0 < \omega_n \le \omega_{\rm cut}} 
 \left\langle \phi^\ast({\bf k}) \left( 
    f +{f^\dagger}^\ast \right) \right\rangle_{\bf k} 
\label{eq:D} 
\end{eqnarray} 
and the vector potential 
\begin{eqnarray}
\nabla\times \left( \nabla \times {\bf A} \right) 
=-\frac{2T}{{{\kappa}}^2}  \sum_{0 < \omega_n} 
 \left\langle {\bf v}
         {\rm Im} g  
 \right\rangle_{\bf k} 
\label{eq:A} 
\end{eqnarray} 
in Eilenberger unit,\cite{KleinJLTP,Ichiokapara}  
with Matsubara frequency $\omega_n$,  
$(g_0N_0)^{-1}=  \ln T +2 T
        \sum_{0 < \omega_n \le \omega_{\rm cut}}\omega_n^{-1} $,   
where 
${\bf v}={\bf k}/k_{\rm F}$ is the direction of Fermi velocity 
${\bf v}_{\rm F}$,  
%and 
${\bf r}$ is the center-of-mass coordinate, 
%$N_0$ is the DOS at the Fermi energy in the normal state, 
and 
$\langle \cdots \rangle_{\bf k}$ indicates the Fermi surface average. 
We use $\omega_{\rm cut}=20 k_{\rm B}T_{\rm c}$.
The internal field 
${\bf B}({\bf r})=\bar{\bf B}+\nabla\times {\bf a}({\bf r})$ 
is related to 
the vector potential 
${\bf A}({\bf r})=\frac{1}{2} \bar{\bf B} \times {\bf r}
 + {\bf a}({\bf r})$ in the symmetric gauge,  
where $\bar{\bf B}=(0,0,\bar{B})$ is a uniform flux density. 
The pairing function is defined as 
$\phi({\bf k})=1 $ for $s$-wave pairing, 
$\phi({\bf k})=\sqrt{2}\cos 2 \theta $ for $d_{x^2-y^2}$-wave pairing. 
In the chiral $p$-wave pairing, we consider two-component 
order parameter 
$\Delta_+({\bf r})\phi_+({\bf k})+\Delta_-({\bf r})\phi_-({\bf k})$
instead of $\Delta({\bf r})\phi({\bf k})$, 
where $\phi_\pm({\bf k})={\rm e}^{\pm{\rm i}\theta}$.\cite{IchiokaP} 
In the $p_+$- ($p_-$-) wave pairing,
$\Delta_+$ ($\Delta_-$) is main component with singular vortex, 
and $\Delta_-$ ($\Delta_+$) is passive component induced around vortices. 

Our calculation is done for $\kappa=2$ and triangular vortex lattice. 
We iterate calculations of eqs. (\ref{eq:eil})-(\ref{eq:A}) 
under given $\bar{B}$, 
and obtain selfconsistent vortex solutions 
for spatial structures of $\Delta({\bf r})$, 
${\bf A}({\bf r})$, and quasiclassical Green's functions, 
as done in previous works.\cite{Ichiokapara,IchiokaP}
Using the solutions, we calculate the external magnetic field $H$ by  
\begin{eqnarray} && 
H=
\bar{B}
+\left\langle \left( B({\bf r})-\bar{B} \right)^2\right\rangle_{\bf r}
/{\bar{B}} 
\nonumber \\ &&   
\hspace{1cm}
+\frac{T}{{\kappa}^2 \bar{B}} 
\sum_{0 < \omega_n} 
\Bigl\langle \Bigl\langle 
\frac{1}{2}{\rm Re}\left\{ 
\frac{(f^\dagger \Delta+f \Delta^\ast)g}{g+1} \right\} 
\nonumber \\ &&   
\hspace{1cm}
+\omega_n {\rm Re}\{ g-1 \} 
\Bigr\rangle_{\bf k} \Bigr\rangle_{\bf r}, 
\label{eq:H}
\end{eqnarray}
which is derived by 
Doria-Gubernatis-Rainer scaling,~\cite{WatanabeKita,Doria,Ichiokapara}  
and $\langle \cdots \rangle_{\bf r}$ indicates the spatial average. 
Magnetic fields are in unit of 
$B_0=\phi_0 /2 \pi R_0^2$ with the flux quantum $\phi_0$ and   
$R_0=\hbar v_{\rm F}/2 \pi k_{\rm B} T_{\rm c}$.  

%%%%%%%%%%%%%%%%%%%%%%%%%%%%
%\section{Results}

%%%%%%%%%%%%%%%%%%%%%%%%%%%%%
\begin{figure}
\begin{center}
\includegraphics[width=8cm]{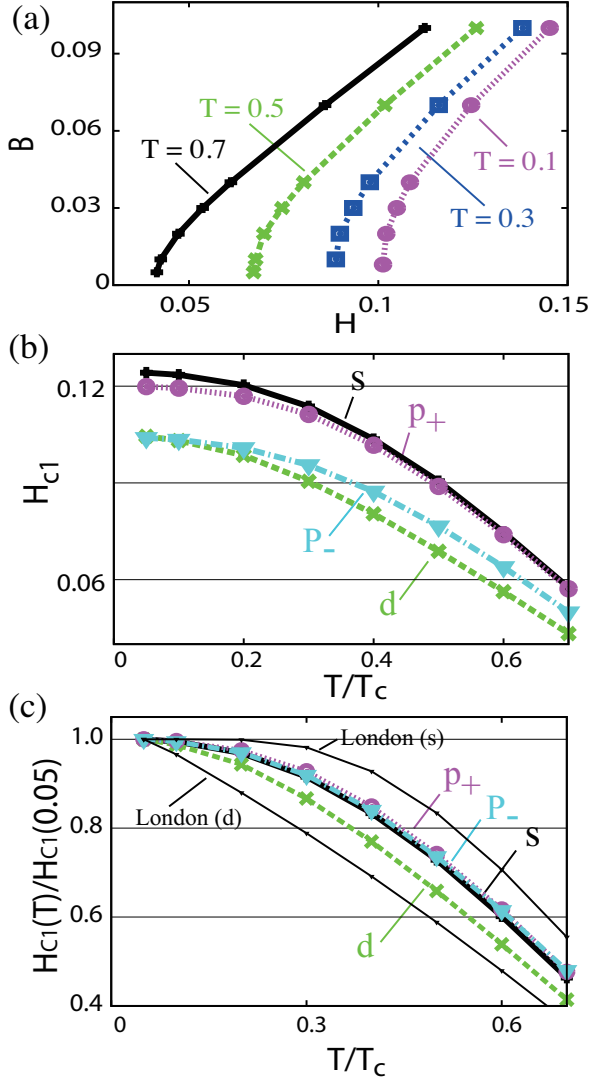}
\end{center}
%\fbox{\rule{0cm}{15cm} \hspace{8cm}}
\caption{
(Color online) 
(a) 
Magnetization curve of $\bar{B}$ as a function of applied field $H$
at $T/T_{\rm c}=0.1$, 0.3, 0.5 and 0.7 for $d$-wave pairing. 
(b) 
$T$-dependence of $H_{\rm c1}(T)$ for $s$-, $d$-, 
$p_+$- and $p_-$-wave pairings, 
estimated by Eilenberger theory. 
(c) 
$H_{\rm c1}(T)$ in (b) is replotted as normalized 
$H_{\rm c1}(T)/H_{\rm c1}(T=0.05T_{\rm c})$. 
We also present normalized $\lambda^{-2}_{\rm London}(T)$ 
for $s$- and $d$-wave pairings. 
}
\label{f1}
\end{figure}
%%%%%%%%%%%%%%%%%%%%%%%%%%%%%

Figure \ref{f1}(a) present magnetization curves of $\bar{B}$ 
as a function of $H$ 
at some $T$ for $d_{x^2-y^2}$-wave pairing. 
There, $H_{\rm c1}$ is defined as onset of $\bar{B}$.   
In Meissner states at $H<H_{\rm c1}$, $\bar{B}=0$.  
In Fig. \ref{f1}(b), we present $H_{c1}(T)$ as a function of $T$ 
for some pairing symmetries,  
and 
we replot them as 
$H_{\rm c1}(T)/H_{\rm c1}(T=0.05T_{c})$ in Fig. \ref{f1}(c) 
to compare the $T$-dependence each other. 

First, we discuss the differences 
between the $s$-wave and the $d$-wave pairings. 
$H_{\rm c1}$ in $d$-wave pairing is smaller than that in $s$-wave pairing,  
because the condensation energy of $d$-wave pairing is weaker 
due to the line node contributions, 
compared to that in the full-gap $s$-wave pairing. 
$H_{\rm c1}$ is related to the energy for creation of a vortex 
in Meissner states.\cite{Fetter}  
As for $T$-dependence, 
reflecting low energy excitations by line nodes,  
$H_{\rm c1}(T)$ in $d$-wave pairing decreases rapidly at low $T$, 
compared with $s$-wave pairing. 

To discuss quantitative validity of the relation 
in eq. (\ref{eq:hc1gl}),  
in Fig. \ref{f1}(c) 
we also present $\lambda_{\rm London}^{-2}(T)$ 
given by
%~\cite{Kogan} 
\begin{eqnarray} 
\lambda_{\rm London}^{-2} 
\propto 
T \sum_{\omega_n} 
\left\langle 
\frac{|\Delta \phi|^2 {\bf v}^2}{(\omega_n^2+|\Delta \phi|^2)^{3/2}} 
\right\rangle_{\bf k}   
\end{eqnarray}
in London theory, 
where $T$-dependence of $\Delta$ is determined by gap eq. (\ref{eq:D}) 
in uniform states. 
In the $s$-wave pairing, as shown in Fig. \ref{f1}(c), 
normalized $H_{\rm c1}(T)$ in Eilenberger theory 
appears smaller than $\lambda_{\rm London}^{-2}(T)$, 
and shows decreases even at low $T$. 
This indicates that the vortex core energy still has $T$-dependence 
at low $T$, rather than saturation expected by $\lambda_{\rm London}(T)$. 
This may include the contribution of vortex core shrink on lowering $T$ 
by Kramer-Pesch effect.~\cite{KP} 
On the other hand, in the $d$-wave pairing, 
$H_{\rm c1}(T)$ in Eilenberger theory is higher than 
$\lambda_{\rm London}^{-2}(T)$. 
Thus, $T$-dependence of the core energy is weaker than estimate by 
$\lambda_{\rm London}(T)$. 
This is an opposite effect to the $s$-wave pairing case, 
and indicates that   
the estimate of core creation energy is not simple 
in $d$-wave pairing because 
we have to consider both contributions inside and outside of vortex cores.  
The latter is contributions by quasiparticles extending toward  
node-directions.\cite{ichiokad}     
These behaviors of $H_{\rm c1}(T)$ is also confirmed for $\kappa=6.9$. 
We expect that the relation in eq. (\ref{eq:hc1gl}) will be examined 
in experiments, comparing $H_{\rm c1}(T)$ with $\lambda^{-2}(T)$. 

Next, we study $H_{\rm c1}(T)$ in chiral $p$-wave superconductors. 
$H_{\rm c1}(T)$ in $p_-$-wave pairing is smaller than 
that in $p_+$-wave pairing. 
This difference in quantitative estimate is consistent to previous results 
by phenomenological GL theory.~\cite{Heeb,Ichioka2005} 
In chiral $p$-wave superconductors, 
opposite chiral component is induced around vortices 
of main chiral component, and 
core energy becomes smaller by the induced component. 
Compared with $p_+$-wave pairing, 
the induced component is larger in $p_-$-wave pairing, and 
the core energy is smaller, making $H_{\rm c1}$ smaller. 
If domains of $p_+$-wave pairing and $p_-$-wave pairing 
coexist at a zero-field, 
on increasing fields vortices penetrate at lower $H_{\rm c1}$ 
only into the $p_-$-wave domain, where chirality is antiparallel to 
the applied field.\cite{Ichioka2005}
As for the $T$-dependence, in Fig. \ref{f1}(c) we see that 
normalized $H_{\rm c1}$ both for $p_+$- and $p_-$-wave pairings 
have similar $T$-dependence to that in $s$-wave pairing. 
This is reasonable, because $p_\pm$-wave pairing with 
$|\phi_\pm|=1$ has full gap, as in $s$-wave pairing. 

%%%

In summary, 
we quantitatively estimated different $T$-dependences of $H_{\rm c1}$ 
between $s$-wave and $d$-wave pairings by Eilenberger theory. 
The $T$-dependences of $H_{\rm c1}(T)$ show 
quantitative deviation from $\lambda_{\rm London}^{-2}(T)$. 
We also studied differences of $H_{\rm c1}(T)$ 
between $p_+$ and $p_-$-wave pairing in chiral $p$-wave superconductors.  
%There, $H_{\rm c1}(T)$ is lower in $p_-$-wave pairing.   
%and shows the same $T$-dependence as in $s$-wave pairing. 
We expect that future experimental studies will confirm the relations of 
$H_{\rm c1}(T)$ and the pairing symmetry in various anisotropic 
superconductors. 

%%%%
We would like to thank K.M. Suzuki and K. Inoue for fruitful discussions, 
and their supports for calculations.

\end{document}